\documentclass[twocolumn,prc,aps,showpacs,superscriptaddress,nofootinbib]{revtex4-1}
\usepackage{amsmath}
\usepackage{amssymb}
\usepackage{graphicx}
\usepackage{color}

\newcommand{\btau}{\mbox{\boldmath $\tau$}} 
\newcommand{\br}{{\bf r}}

\usepackage{ulem}

\begin{document}
\title{The $N$=50 and $Z$=28 shell closure revisited}
\author{T. R. Routray}
\email{trr1@rediffmail.com}
\affiliation{School of Physics, Sambalpur University, Jyotivihar-768 019, India.}
\author{P. Bano}
\email{mailme7parveen@gmail.com}
\affiliation{School of Physics, Sambalpur University, Jyotivihar-768 019, India.}
\author{M. Anguiano}
\email{mangui@ugr.es}
\affiliation{Departamento de F\'{\i}sica At\'omica, Molecular y Nuclear,
Universidad de Granada, E-18071 Granada, Spain}
\author{M. Centelles}
\email{mariocentelles@ub.edu}
\affiliation{Departament de F\'isica Qu\`antica i Astrof\'isica and Institut de Ci\`encies del Cosmos (ICCUB),
Facultat de F\'isica, Universitat de Barcelona, Mart\'i i Franqu\`es 1, E-08028 Barcelona, Spain}
\author{X. Vi\~nas}
\email{xavier@fqa.ub.edu}
\affiliation{Departament de F\'isica Qu\`antica i Astrof\'isica and Institut de Ci\`encies del Cosmos (ICCUB),
Facultat de F\'isica, Universitat de Barcelona, Mart\'i i Franqu\`es 1, E-08028 Barcelona, Spain}
\author{L.M. Robledo}
\email{luis.robledo@uam.es}
\affiliation{Departamento de F\'isica Te\'orica and CIAFF,
Universidad Aut\'onoma de Madrid, E-28049 Madrid, Spain}
\affiliation{Center for Computational Simulation,
Universidad Polit\'ecnica de Madrid,
Campus de Montegancedo, Boadilla del Monte, E-28660 Madrid, Spain}
\date{\today}

\def \beq {\begin{equation}}
\def \eeq {\end{equation}}
\def \beqa {\begin{eqnarray}}
\def \eeqa {\end{eqnarray}}
\def \lf {\left}
\def \ri {\right}
\newcommand{\mbold}[1]{\mbox{\boldmath$#1$}}
\def \ep {{\cal E}}
\def \kf {k_{\rm F}}
\def \wt {\widetilde}
\def \Hw {H_{\rm w}}
\def \scdot {\!\cdot\!}
\def \bnabla {\mbold{\nabla}}
\def \deltap {\delta^\prime}
\def \deltapp {\delta^{\prime\prime}}

\begin{abstract}
Recent experiments performed in neutron-rich copper isotopes have revealed a crossing in the nucleus $^{75}$Cu between the $3/2^-$ and $5/2^-$ 
levels, which correspond to  the ground-state and the first excited state in isotopes with mass number below $A =75$. Due 
to the strong single-particle character of these states, this scenario can be investigated through the analysis of the proton spectrum provided 
by mean-field models in nickel isotopes with neutron numbers between $N$=40 and $N$=50. In this work we show that the aforementioned crossing 
is mainly driven by the mean-field provided by the effective nucleon-nucleon and spin-orbit interactions. We also analyze the impact of the tensor 
interaction, and find that in some mean-field models it is essential to reproduce the crossing of the 2$p_{3/2}$ and 1$f_{5/2}$ proton 
single-particle levels, as in the case of the SAMi-T Skyrme force and the D1M Gogny interaction, whereas in other cases, as for example the SLy5 
Skyrme force, a reasonable tensor force appears to be unable to modify the mean-field enough to reproduce this level crossing. 
Finally, in the calculations performed with the so-called simple effective interaction (SEI), it is shown that the experimental data in nickel
and copper isotopes considered in this work can be explained satisfactorily without any explicit consideration of the tensor interaction. 
\end{abstract} 

\pacs{}
\keywords{}

\maketitle

Effective nuclear mean field models are usually fitted to reproduce as well as possible the
structure and properties of nuclei along the stability valley. However, modern facilities, such
as SPIRAL at GANIL, the Radioactive Beam Experiment at ISOLDE at CERN, the Facility for Antiprotons 
and Ion Research (FAIR) at GSI and the Facility for Rare Isotope Beams (FRIB) at MSU, set to be
in full operation in 2022, are delivering many new experimental data about exotic nuclei near the drip lines. Of particular 
relevance  are the changes observed in the nuclear shell structure where new phenomena, such as the 
disappearance of the standard magic numbers 20 and 28 for neutrons and the emergence of new 
magic numbers at $N$=14, 16, 32 and 34, may occur in neutron-rich nuclei (see \cite{olivier17}
 and references therein). These new magic numbers appear due to the imbalance of 
neutron and protons, which strongly modify the spin-orbit potential that in turn determines 
the shell structure.

A region of experimental interest nowadays is around the magic numbers $Z$=28 and $N$=50, where 
measurements of the decay properties in Co, Ni, Cu and Zn reveal the magic character of the
nucleus $^{78}$Ni \cite{franchoo98,xu14,sahin17}, which is also confirmed by the precise 
measure of the nuclear spin and dipole and quadrupole moments in neutron-rich Cu-isotopes 
\cite{flanagan09,groote17} and by the $\gamma$-ray spectroscopy in $^{79}$Cu \cite{olivier17}. 
The experimental results also show an inversion of the spin of the ground-state of 
neutron-rich copper isotopes from 3/2$^{-}$ to 5/2$^{-}$, which takes place beyond $^{73}$Cu 
\cite{stefanescu08}. 
Large scale shell-model calculations reported in Ref.\cite{smirnova04} show 
that in copper isotopes with neutron number larger than $N=40$ the 3/2$^{-}$ and 5/2$^{-}$ 
states have an important single-particle character, although coexisting with other states of 
collective nature. More recent shell-model calculations \cite{nowacki16} also predict the 
double-magic character of the nucleus  $^{78}$Ni, which, however, shows the phenomenon of shape coexistence. 

At mean field level the crossing of the 2$p_{3/2}$ and 1$f_{5/2}$ single-particle proton levels in 
neutron-rich Ni isotopes can be understood in terms of a strong tensor interaction, which is 
attractive between the neutron 1$g_{9/2}$ and proton 1$f_{5/2}$ levels and 
repulsive between 1$g_{9/2}$ and 2$p_{3/2}$ levels \cite{otsuka05a}. The microscopic nucleon-nucleon 
interaction contains 
a tensor contribution, whose most important component is related to the tensor-isospin channel.
In this case the long-range behavior of the tensor force is dominated by the exchange of a 
single pion \cite{machleidt87,wiringa95}, which is longer
than the ranges of all the remaining contributions to the nucleon-nucleon interaction.     

\begin{figure}[t]
\includegraphics[width=8cm]{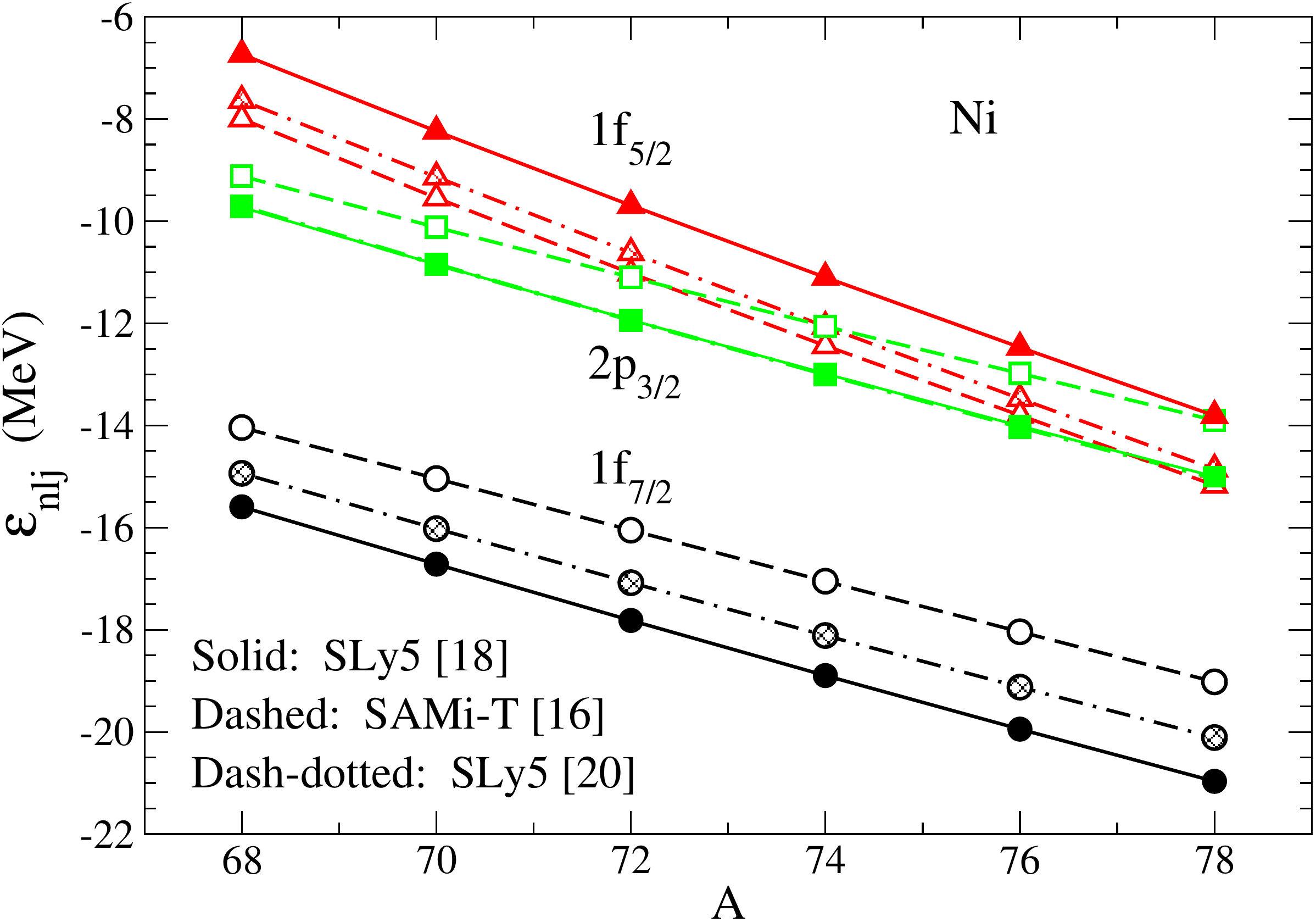}
  \caption{Proton single-particle levels around the Fermi level for Ni isotopes from $A$=68 to 
$A$=78 computed with the Skyrme forces SAMi-T \cite{shen19} and SLy5 with the tensor part fitted in 
Refs.~\cite{colo07} and \cite{grasso13}, respectively.} 
  \label{figure1}
  \end{figure}

In the case of Skyrme forces a tensor interaction, which is zero-range and momentum dependent
in configuration space, was proposed long ago \cite{stancu77}. 
The question raised on the validity of the use of the zero-range tensor force in Ref.~\cite{otsuka05b}
is clarified later on in the recent work by Brink and Stancu \cite{brink18}, where they have shown that 
the momentum dependence of the zero-range tensor force \cite{stancu77} simulates the same effect as the 
finite-range tensor interaction. With Skyrme forces, the use of a tensor 
term seems unavoidable to reproduce the position of the 2$p_{3/2}$ and 1$f_{5/2}$ single-particle 
proton levels in the Ni isotopic chain beyond neutron number $N$=40 \cite{brink18} 
observed experimentally in Refs.~\cite{franchoo98,stefanescu08,flanagan09,sahin17} and interpreted
theoretically with  Monte-Carlo shell model calculations \cite{smirnova04,nowacki16}. In Fig.
\ref{figure1} we show the single-particle energies of the 1$f_{7/2}$, 2$p_{3/2}$ and 1$f_{5/2}$
proton levels along the Ni isotopic chain with neutron number in the range 
$N$=40-50 obtained with the Skyrme interaction SAMi-T that includes tensor terms \cite{shen19}. This force predicts the 
crossing between the 2$p_{3/2}$ and 1$f_{5/2}$ proton levels in the nucleus $^{72}$Ni.
However, it should be pointed out that the incorporation of tensor terms in the Skyrme interaction may 
not be sufficient to reproduce the crossing of the 2$p_{3/2}$ and 1$f_{5/2}$ single-particle 
proton levels in neutron-rich Ni isotopes. For example, this appears to be the case of the Skyrme interaction SLy5 
\cite{chabanat98} including tensor terms, which fails to reproduce the proton 
level crossing in Ni isotopes as can be seen in Fig.~\ref{figure1}. This is surprising 
as this force was fitted to reproduce the relative energies of 
proton (neutron) single-particle levels in some isotopic (isotonic) chains \cite{colo07} 
determined experimentally in Ref.~\cite{schiffer04}. A reparametrization 
of the tensor part of the SLy5 interaction has been proposed recently \cite{grasso13} by fitting the 
the neutron 1$f$ spin-orbit splitting for the nuclei $^{40}$Ca, $^{48}$Ca and 
$^{56}$Ni to the experimental data. Again, this new parametrization of the 
tensor force does not predict the crossing of the aforementioned levels, which can be seen 
in the same figure.

\begin{figure}[t]
  \includegraphics[width=8cm]{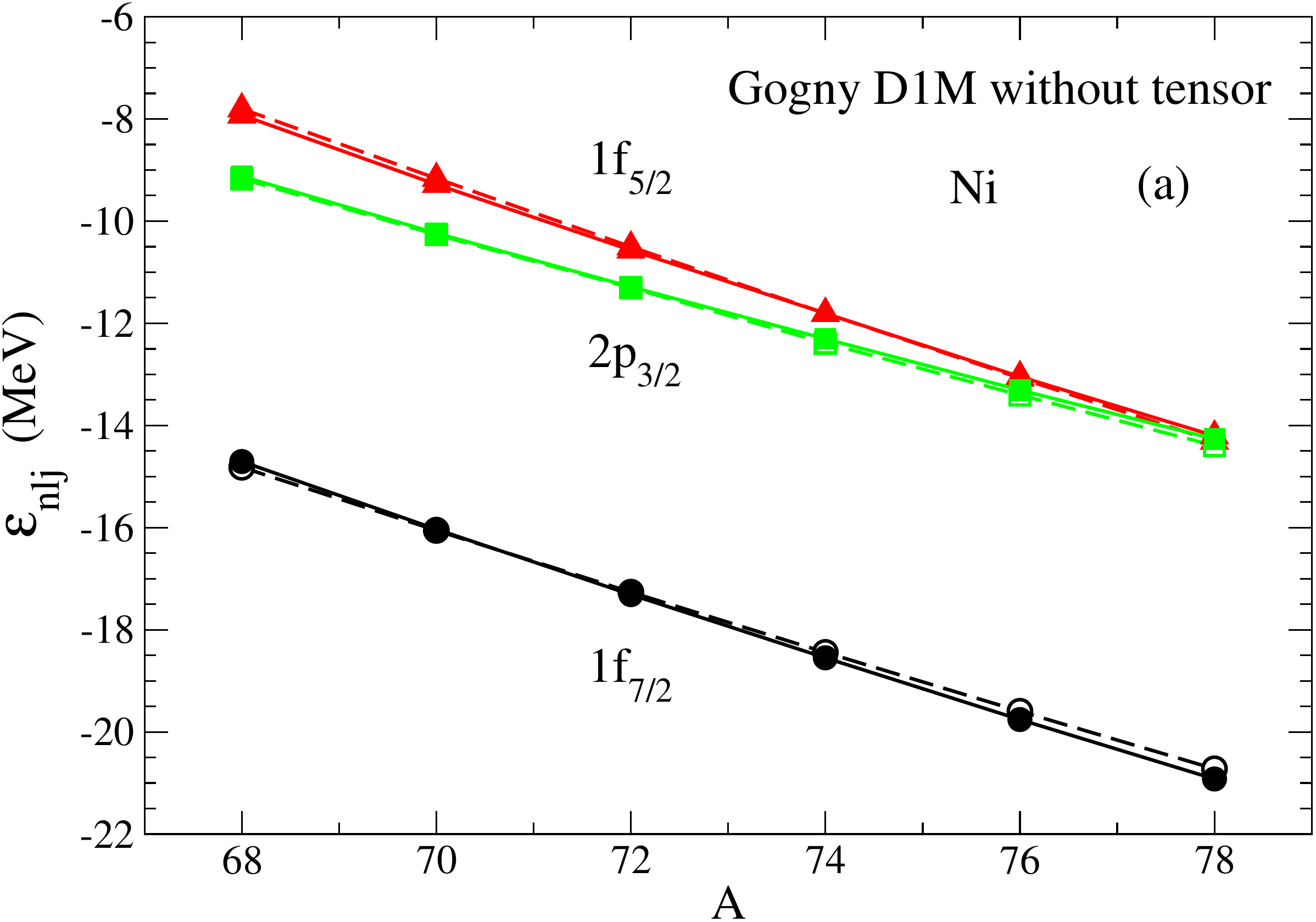}
  \includegraphics[width=10cm]{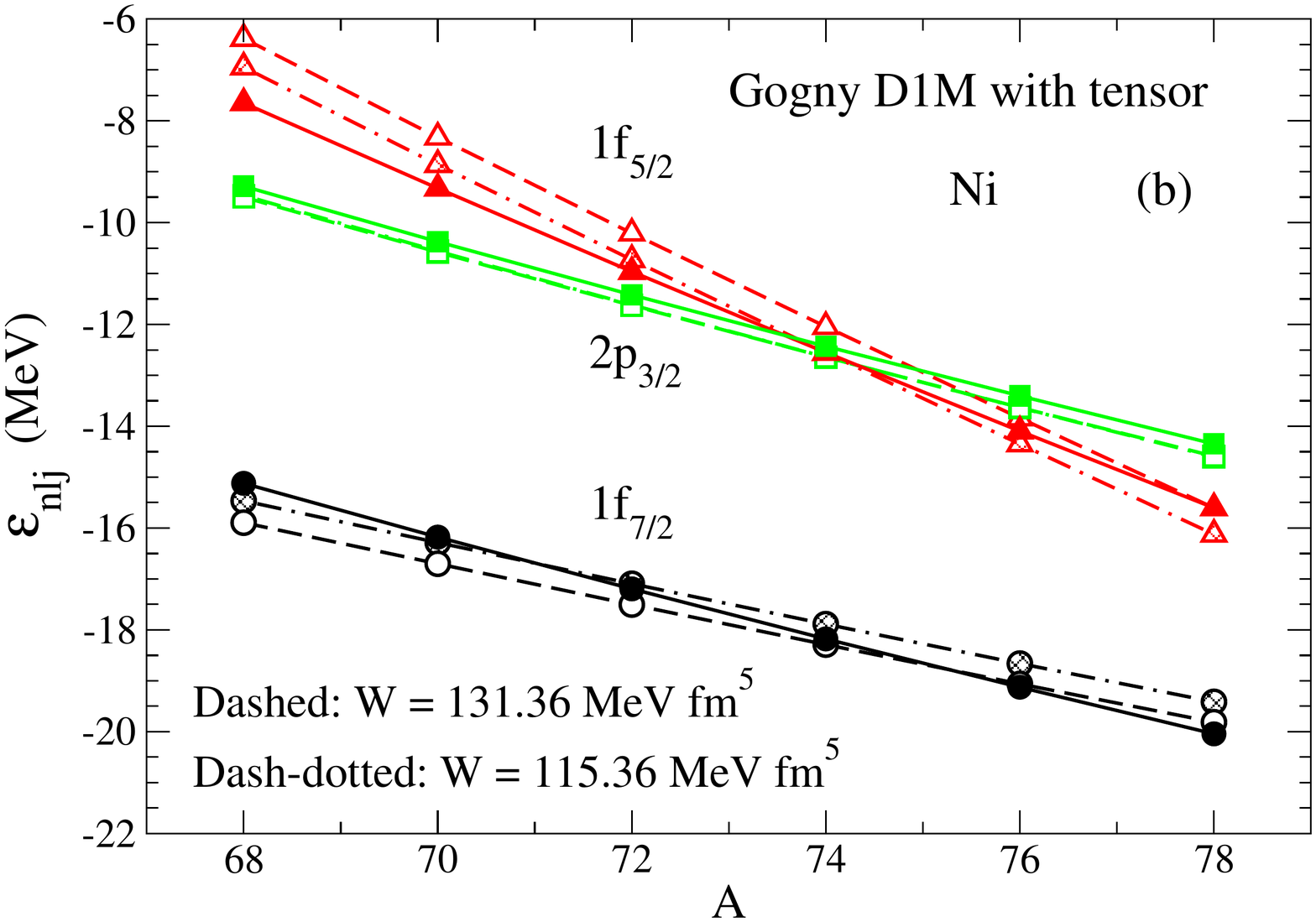}
  \caption{Proton single-particle levels around the Fermi level for Ni isotopes in the range of mass numbers between $A$=68
and $A$=78 computed with the D1M Gogny force without (upper panel) and with (lower panel) tensor terms.
Solid lines are obtained from the HF calculation of Ref.~ \cite{anguiano12} using the D1MTd interaction. Dashed and dashed-dotted 
lines are computed within the QLDFT approximation described in Ref.~\cite{soubbotin03}.}
  \label{figure2}
  \end{figure}

The standard parametrizations of the Gogny force \cite{decharge80} of the D1 family, namely 
D1S \cite{berger91}, D1N \cite{chappert08} and D1M \cite{goriely09}, predict that, for increasing 
neutron number, the energy gap between the 2$p_{3/2}$ and 1$f_{5/2}$ single-particle proton levels in 
neutron-rich Ni isotopes decreases  and almost vanishes in the magic nucleus $^{78}$Ni. This can be seen 
in the upper panel of Fig.~\ref{figure2} that displays the single-particle energies of the proton levels 
1$f_{7/2}$, 2$p_{3/2}$ and 1$f_{5/2}$  of the nuclei between $^{68}$Ni and $^{78}$Ni 
computed with the D1M interaction. In this panel we show the full Hartree-Fock (HF) result 
calculated in coordinate space (see \cite{anguiano16} and references therein) and the one 
obtained using the so-called Quasilocal Density Functional Theory (QLDFT) \cite{soubbotin03},
where the exchange energy is written as a local density functional  with the help of the semi-classical 
one-body density matrix including $\hbar^2$ corrections \cite{soubbotin00}. From this figure we can conclude 
that the QLDFT single-particles energies agree very accurately with the full HF ones.
If a tensor contribution is included in the Gogny force \cite{anguiano11,anguiano12,anguiano16},
the HF+BCS calculation performed with the D1MTd interaction in neutron-rich Ni isotopes predicts 
the same eigenvalues for the 2$p_{3/2}$  and 1$f_{5/2}$ proton levels in the nucleus $^{74}$Ni. 
The same result can be obtained using QLDFT if a tensor contribution is included. Unlike the calculations 
reported in Refs.~\cite{anguiano11,anguiano12,anguiano16}, we include in the 
QLDFT functional a zero-range tensor force as the one used in Skyrme interactions \cite{stancu77}.
The predictions of the HF+BCS calculation using the D1MTd parametrization are 
compared in the lower panel of Fig.~\ref{figure2} with those provided by the QLDFT with D1M
supplemented by a zero-range tensor force with parameters $\alpha_T$=$-$52.08 MeV fm$^5$ and 
$\beta_T$=215.83 MeV fm$^5$ (see \cite{grasso13,brink18} for the definition of these parameters). 
The tensor term used for the D1MTd interaction has been defined as:
\begin{eqnarray}
V_{\rm T}(1,2) &=& \left[ V_{\rm T1} + 
V_{\rm T2} \frac{\left ( 1\, + \,  \btau(1) \cdot \btau(2) \right )}{2}
\right] S_{12} 
\nonumber\\[1mm] \mbox{}
&& \times \exp{\left [-(\br_1 - \br_2)^2 / \mu_{\rm T}^2 \right ] }  \, ,
\label{eq:tensor}
\end{eqnarray}
where $S_{12}$ is the traditional tensor operator and $\btau (i)$ the isospin operator. The specific values of the parameters are 
$V_{\rm T1}=-230$ MeV,  $V_{\rm T2}= 180$  MeV and $\mu_{\rm T} = 1.0$ fm, and have been chosen to reproduce the energy values of 
the first $0^-$ states for the nuclei $^{16}$O, $^{40}$Ca and $^{48}$Ca. 
The nice agreement between both results  points out, in agreement with Ref.~\cite{brink18}, that
the main effects of the tensor interaction can be very well reproduced by using a zero-range tensor force
similar to that associated to the Skyrme forces \cite{stancu77}. It should be pointed out that when the tensor force is
added in the QLDFT functional the parameters of the interaction should be readjusted. Here, in order to estimate the impact of the additional 
tensor force on the binding energies and single-particle spectra in a simple way, we refit the strength of the spin-orbit force to keep the 
binding energy of the nucleus $^{208}$Pb equal to the original value predicted by the D1M force. This implies that the spin-orbit strength changes 
from the original D1M value 115.36 MeV fm$^5$ to 131.36 MeV fm$^5$ when the zero-range tensor force is included in the QLDFT functional. With 
this change the binding energy of the nucleus $^{78}$Ni increases by almost 1.5\%,  while the crossing of the 2$p_{3/2}$  and 1$f_{5/2}$ levels is 
slightly modified but still predicting the crossing of the two first proton single-particle excited states in passing from $^{74}$Ni to $^{76}$Ni.  

In this work we would like to note the fact that, in spite of the results discussed till now, the tensor 
interaction may not be  necessary to reproduce the crossing between the 2$p_{3/2}$ and 1$f_{5/2}$ single-particle 
proton levels in neutron-rich Ni isotopes. 
This is the case, for instance, with the so-called simple effective interaction (SEI), which was first 
proposed for symmetric and asymmetric nuclear matter studies in Ref. \cite{behera98} and extended later on 
to finite nuclei in Refs.~\cite{behera13,behera15}. This effective interaction contains a single 
finite-range term with a form factor $f(r)$ of Gauss or Yukawa type and two zero-range terms, one of them 
density-dependent with an additional factor to avoid a supra-luminous behavior \cite{behera97}.
Thus, the SEI reads 
\begin{eqnarray}\label{eq1}
V_{\rm eff} &=& t_{0}(1+x_{0}P_{\sigma})\delta(\vec{r}) \nonumber\\[1mm]
&& \mbox{} +\frac{t_{3}}{6}(1+x_{3}P_{\sigma})\left(\frac{\rho(\vec{R})}
{1+b\rho(\vec{R})}\right)^{\!\!\gamma}\delta(\vec{r})
\nonumber \\[1mm]
&& \mbox{}
+(W+BP_{\sigma}-HP_{\tau}-MP_{\sigma}P_{\tau})f(\vec{r}) \,.
\end{eqnarray}
The fitting protocol of this interaction, explained in detail in Refs.~\cite{behera98,behera13,behera15}, differs in several 
respects from the one used to obtain the parameters in Skyrme, Gogny or M3Y forces. Nine of the eleven parameters of SEI are 
fitted to reproduce empirical constraints and microscopical results in nuclear and neutron matter obtained with realistic
interactions. In particular it is demanded that the nuclear mean-field in symmetric nuclear matter at saturation density 
vanishes for a kinetic energy of the incident nucleon of 300 MeV, a value extracted from the optical model 
fit to the nucleon-nucleus scattering data at intermediate energies.  
This constraint allows to determine, for a given value of the exponent $\gamma$, the strength of the exchange 
energy and the range of the form-factor in an unambiguous way. One of the two free parameters, namely $x_0$, is fixed 
from the spin-up spin-down splitting of the effective mass in polarized neutron matter \cite{behera15}. Finite nuclei 
calculations require, in addition, to consider the spin-orbit interaction, which is chosen of zero-range as in the case of 
Skyrme or Gogny forces \cite{vautherin72}. The $t_0$  parameter of SEI and the strength of the spin-orbit interaction $W_0$ 
are fitted within the QLDFT to reproduce the binding energies of the magic nuclei $^{40}$Ca and $^{208}$Pb. 
To deal with open-shell nuclei we introduce a density-dependent pairing force proposed by Bertsch and Ebensen \cite{bertsch91} 
without any adjustable parameter and treat the pairing correlations within an improved BCS approach \cite{delestal01} 
(see Refs.~\cite{behera13,behera15} for further details). In Ref.~\cite{behera13} we have analyzed the binding energies 
of 161 even-even spherical nuclei using the QLDFT formalism. 
This study was enlarged to 620 even-even spherical and deformed nuclei described at HFB level using SEI
\cite{behera16}. These studies performed in finite nuclei show that, on the one hand, the binding energies obtained in 
the QLDFT approximation with this parametrization of SEI are in excellent agreement with the corresponding full HFB results and, 
on the other hand, that the {\it rms} deviations predicted by this set for the binding energies and charge radii of 620 even-even nuclei are similar to those found using effective forces of Skyrme and Gogny types. It is also to be pointed out that the single-particle 
energies computed with the SEI parametrization used in this work are in reasonable agreement with the experimental values and 
describe the spectrum of $^{208}$Pb as good or better than other effective mean-field models (see in this respect Fig.~8 of 
\cite{behera13}).     
The mean-field predicted by SEI is also able to predict, without any additional modification, the kinks in the isotopic shifts 
of charge radii in $^{208}$Pb, $^{210}$Pb and $^{212}$Pb, which are not predicted by Skyrme or Gogny forces with an isospin-independent 
spin-orbit interaction (see \cite{behera13} for a more detailed discussion). 

\begin{figure}[t]
  \includegraphics[width=10cm]{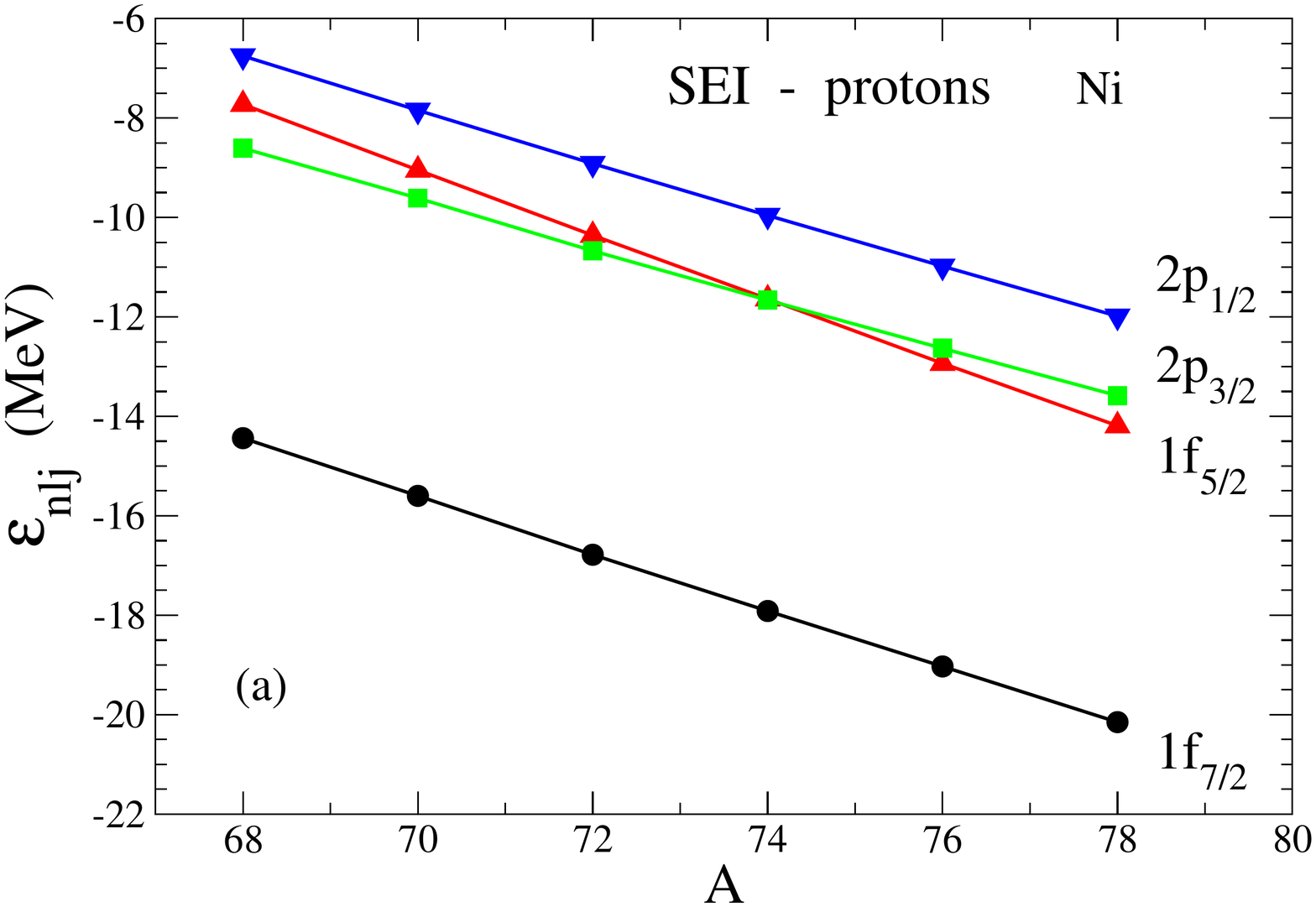}
  \includegraphics[width=10cm]{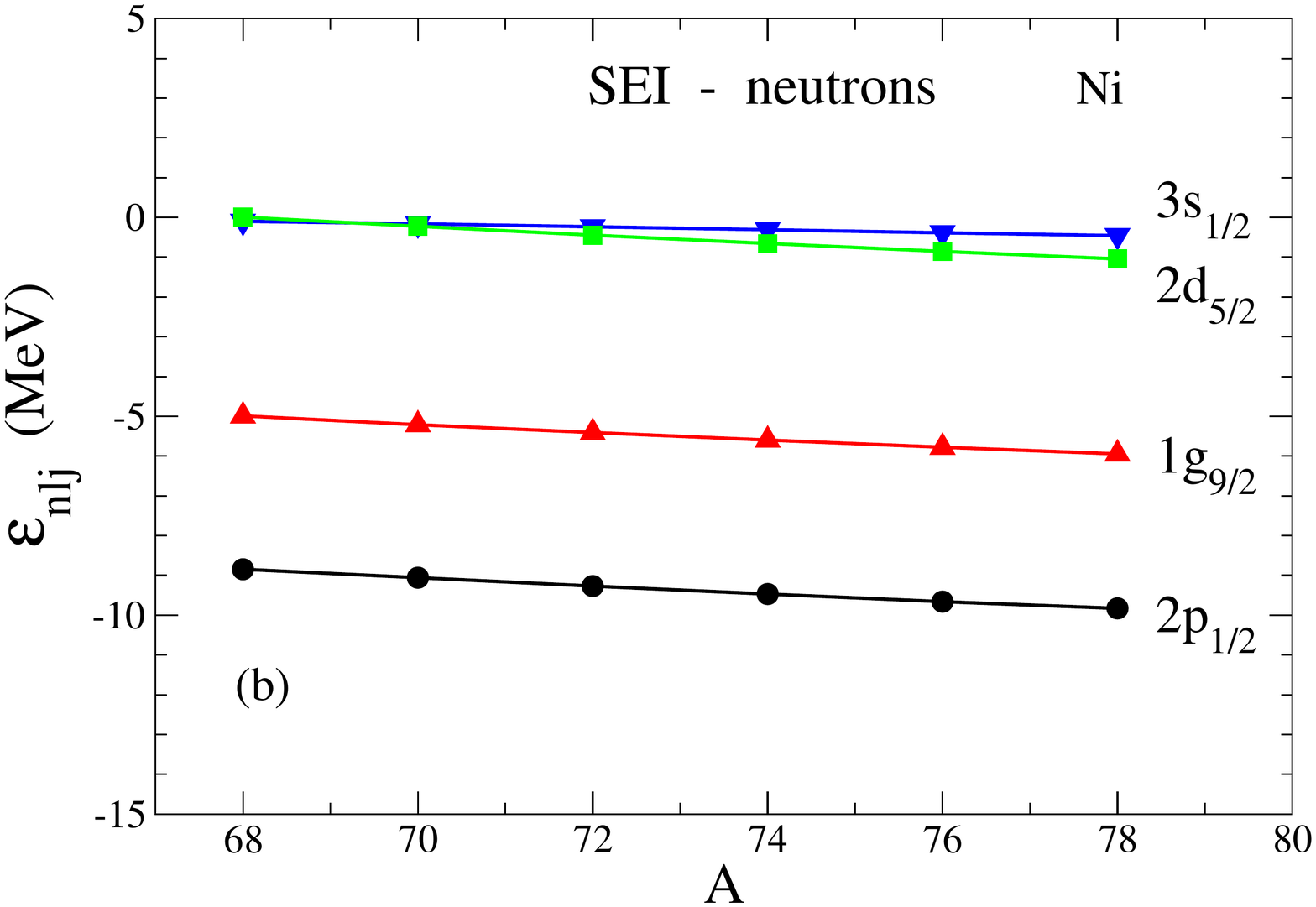}
  \caption{Proton (upper panel) and neutron (lower panel) single-particle levels around the Fermi level for Ni isotopes in the range 
of mass numbers between $A$=68 and $A$=78 predicted by the SEI model discussed in Ref.~\cite{behera13}.}
  \label{figure3}
  \end{figure}

In Fig.~\ref{figure3} we display the proton (upper panel) and neutron (bottom panel) single-particle levels around 
the Fermi energy in neutron-rich Ni isotopes with neutron number from $N$=40 to $N$=50 computed using the parametrization 
of Ref.~\cite{behera13}. As can be seen in the upper 
panel the splitting between the 1$f_{5/2}$ and 1$f_{7/2}$ proton levels decreases in passing from $N$=40 to $N$=50 by 
0.75 MeV, which is about one-half of the proton gap reduction found in large-scale shell-model calculations \cite{welker17}. 
This reduction is due to the strong 1$f_{5/2}$-1$g_{9/2}$ proton-neutron attraction combined with the 1$f_{7/2}$-1$g_{9/2}$ 
repulsion, which increases when the occupation of the 1$g_{9/2}$ level grows \cite{otsuka05a,otsuka01,otsuka10}. 
A similar effect, but less pronounced, also happens with the 2$p_{3/2}$-2$p_{1/2}$ proton gap,
which is reduced by 0.26 MeV in passing from $^{68}$Ni to $^{78}$Ni. As explained in detail in 
Refs.~\cite{otsuka01,otsuka05a,otsuka10}, these changes in the single-particle energies are due to the monopole contributions 
of the original interaction, which includes central, spin-orbit and eventually tensor contributions.
As a result of the mentioned evolution of the Ni proton levels with neutron number in SEI, this interaction leads
to the crossing between the 2$p_{3/2}$ and 1$f_{5/2}$ proton levels in neutron-rich Ni isotopes at $N=46$, in agreement with 
the experimental observation.

The experimental results in Cu isotopes suggest that the crossing between the 2$p_{3/2}$ and 1$f_{5/2}$ proton levels take 
place in the nucleus $^{75}$Cu, which implies that the ground-state of $^{79}$Cu has spin-parity 5/2$^-$ \cite{olivier17}. 
In the same nucleus it is also found that the first excited state has spin-parity 3/2$^-$ and lies 656 keV above the ground 
state. The nature of the low-lying levels in Cu isotopes with neutron numbers beyond $N$=40 have been investigated by means 
of Coulomb excitation with radioactive beams \cite{stefanescu08}. These experimental results reveal that at very low excitation 
energy collective and single-particle levels coexist. 
The  3/2$^-$ and  5/2$^-$ levels are of particular interest and their excitation energies can be easily 
estimated from the single-particle energies computed with a mean-field model. The QLDFT calculation of $^{78}$Ni with the 
SEI model without adding any extra tensor term predicts that the excitation energy of the 3/2$^-$ level is 607 keV, to be 
compared with the value 656 keV, extracted from the analysis of experimental data \cite{olivier17},  and 294 keV obtained 
in the shell-model calculations reported in the same reference. The analysis of the experimental data of $^{79}$Cu also 
suggests another excited state 1/2$^-$ 1511 keV above its ground state, while the excitation energy is 1957 keV according to 
shell-model calculations and it is predicted at 2203 keV by the SEI calculation in $^{78}$Ni. 
The level structure of the nucleus $^{77}$Cu has been investigated in Ref.~\cite{sahin17}. The ground-state is also 5/2$^-$ 
and the first excited state 3/2$^-$ lies 293 keV above the ground-state while the shell-model prediction is 184 keV. This 
experimental scenario is also described, at least qualitatively, by the SEI calculations carried out in $^{76}$Ni, where the 
ground-state is predicted to be 5/2$^-$ and the first excited state, 3/2$^-$, is placed 301 keV above the ground state.     
  
A slightly more accurate estimate about the single-particle properties of neutron-rich Cu isotopes can be obtained by 
performing mean-field calculations of odd $^{69}$Cu-$^{75}$Cu isotopes with blocking in the uniform filling approximation \cite{Perez08}. 
Although these nuclei are slightly deformed \cite{hilaire07}, we neglect the deformation 
effects in our estimate of the energies of the ground-state and the first excited state. The results of our investigations 
using the SEI model are collected in Table \ref{table1} together with the experimental energies extracted from Figure 3 of 
\cite{olivier17}. The SEI model predicts that the crossing between the 3/2$^-$ and 5/2$^-$ single-particle states occurs for 
the nucleus $^{75}$Cu, in agreement with the experimental findings. The excitation energies of the first excited level 
5/2$^-$, in isotopes between $^{69}$Cu and $^{75}$Cu, and 3/2$^-$, for  $^{77}$Cu and $^{79}$Cu, predicted by the SEI model 
are in nice agreement with the experimental data. We have repeated this analysis but using the D1M and D1MTd Gogny   
forces in full HF calculations. We obtain that using the D1MTd interaction, the experimental spin-parity of the ground-state is predicted
correctly and the experimental energy of the first excited state is qualitatively reproduced (see the column of results for D1MTd in 
Table \ref{table1}). On the other hand, using the D1M force the crossing of the ground and first excited states is not predicted and 
the energy of the first excited state follows a downwards trend in disagreement with the experimental results. 
\begin{table*}[thb]
\begin{center}
\caption{Ground-state spin and energy of neutron-rich odd Cu isotopes predicted by the SEI model used in this work. The energy of the 
first excited state  $E^*$ is shown for the SEI model and for the HF calculation in a simple IPM approximation using the D1M and D1MTd 
Gogny forces. Notice that D1M predicts 3/2$^-$ as spin-parity of the ground-state of the nuclei $^{77}$Cu and $^{79}$Cu. 
The experimental energies are also reported for comparison. Also notice that according to the experimental results of Ref.~\cite{flanagan09}, 
the spin-parity of the ground-state of the nucleus $^{75}$Cu is 5/2$^-$ and the first excited state 3/2$^-$ 62 keV above.} \vspace{5pt}
\label{table1}
\begin{tabular}{c|ccccccc} \hline
 Nucleus           &   Spin-Parity  & Energy(SEI) & Energy(exp) & $E^*$(SEI) &  $E^*$(exp) &   $E^*$(D1M)  &  $E^*$(D1MTd) \\
                   &                & (MeV)       & (MeV)       & (keV)      & (keV)        & (keV)        & (keV) \\ \hline
   $^{69}$Cu       &      3/2$^-$     & $-$598.59  & $-$599.97 & 794    &   1215    & 1199  &  1635   \\
   $^{71}$Cu       &      3/2$^-$     & $-$612.93  & $-$613.09 & 544    &    537    &   952  &   1048  \\
   $^{73}$Cu       &      3/2$^-$     & $-$625.76  & $-$625.51 & 282    &    263    &   719  &   458  \\
   $^{75}$Cu       &      3/2$^-$     & $-$637.49  & $-$637.13 &  72    &     62    &   499  &   123   \\
   $^{77}$Cu       &      5/2$^-$     & $-$648.38  & $-$647.42 & 246    &    295    & 264    &   692   \\
   $^{79}$Cu       &      5/2$^-$     & $-$658.19  & $-$656.65 & 525    &    660    &  61    &   1257   \\
\hline
\end{tabular}
\end{center}
\end{table*}

There is not much experimental information available concerning the neutron single-particle levels in neutron-rich Ni 
isotopes between $^{68}$Ni and $^{78}$Ni \cite{brink18}. The SEI predicts that the gap between the 1$g_{9/2}$ and 
2$d_{5/2}$ neutron levels remains practically constant for all the neutron-rich isotopes considered, pointing out that 
the SEI model maintains the magic character of the neutron number $N$=50. This fact is in agreement with a similar finding 
reported in Ref.~\cite{brink18}, which for the case of the Skyrme III interaction including a tensor contribution, predicts 
an enhancement of the neutron gap between the 1$g_{9/2}$ and 2$d_{5/2}$ levels in passing from $^{68}$Ni to $^{78}$Ni, 
reinforcing the magic character of the neutron number $N$=50.

In summary, in this work we wanted to underline that the effect of the monopole component of the central 
part of the nucleon-nucleon interaction is actually relevant and may modify the behavior 
of the single-particle levels along isotopic or isotonic chains in a quite considerable 
extension, masking the monopole effects coming from the spin-orbit and tensor parts of the 
nucleon-nucleon interaction. To highlight these facts we consider exotic nickel isotopes 
between $^{68}$Ni and $^{78}$Ni because different measurements performed in copper isotopes 
suggest a crossing between the unoccupied 2$p_{3/2}$ and 1$f_{5/2}$ single-particle proton 
levels that occurs when the neutron number is $N$=46. It has been claimed in earlier 
literature \cite{brink18} that using Skyrme forces including a zero-range tensor term the 
aforementioned crossing can be reproduced. This is true for some particular forces, as for 
example the Skyrme III interaction used in \cite{brink18} or in the recently reported SAMi-T 
force \cite{shen19}. However, in the particular case of the SLy5 interaction \cite{chabanat98}
the crossing between the 2$p_{3/2}$ and 1$f_{5/2}$ proton levels does not seem to be reproduced by adding 
tensor terms \cite{colo07,grasso13}, which were fitted to reproduce other observables 
sensitive to the tensor force but not to the crossing of proton levels in neutron-rich Ni isotopes.
 This is a first indication that the monopole effects from the tensor force may not be enough to 
reproduce the aforementioned crossing. It also suggests that in the case of the SLy5 interaction the 
central and spin-orbit components of the mean field are not well suited to reproduce the crossing of the 
proton levels in neutron rich Ni isotopes. Another interesting example is provided 
by the finite-range D1M Gogny interaction \cite{goriely09}. This Gogny force, as well as the 
D1S one \cite{berger91} widely used as a benchmark for pairing and deformation properties in 
finite nuclei \cite{hilaire07}, predicts that the single-particle energies of the 2$p_{3/2}$
 and 1$f_{5/2}$ proton levels almost coincide for the nucleus $^{78}$Ni, 
implying that the monopole contribution of the central and spin-orbit parts of the 
nucleon-nucleon interaction are not enough for yielding the crossing in the right place. 
However, the right crossing can be achieved by adding a tensor force 
\cite{anguiano16,anguiano11,anguiano12}. We have repeated this study of neutron-rich nickel 
isotopes described by the D1M Gogny force by using the QLDFT formalism instead of HF in 
coordinate space. We have seen that the QLDFT energy levels coincide almost perfectly with the 
ones obtained at HF level, at least  for the Ni isotopes analyzed here.    
We have included in the QLDFT formalism a zero-range tensor term as the one used in 
Skyrme interactions \cite{stancu77}, finding again a very good agreement with the HF calculation 
that includes a finite-range tensor force.    
As a last example, we discuss the predictions of QLDFT calculations obtained with the SEI 
interaction used in Refs.~\cite{behera13,behera15}, which does not contain tensor terms. In this case, the crossing of 
the 2$p_{3/2}$ and 1$f_{5/2}$ proton levels in the nickel isotopic chain takes place at the 
nucleus $^{74}$Ni without any modification of the interaction and its parameters. We have refined the SEI predictions 
by performing mean-field calculations along the $^{69}$Cu-$^{79}$Cu
isotopic chain including blocking in the uniform filling approach and neglecting deformation 
effects. These calculations predict that the ground-state and the first excited state are 
3/2$^-$ and 5/2$^-$, respectively, for isotopes lighter than $^{75}$Cu where the crossing takes
 place, in agreement with the compilation of experimental data of Fig.~3 of 
Ref.~\cite{olivier17}. The spin-parity of the ground state and first excited state of the 
nuclei $^{77}$Cu and $^{79}$Cu are predicted to be 5/2$^-$ and 3/2$^-$, respectively, also in 
agreement with the experimental data. In addition our estimate also reproduces the energies of 
the first excited state of copper isotopes in the range $^{69}$Cu-$^{79}$Cu to quite satisfactory extent.

\begin{acknowledgments}
M.C. and X.V. were partially supported by Grants No.\ FIS2017-87534-P from MINECO and FEDER
and No.\ CEX2019-000918-M from the State Agency for Research of the Spanish
Ministry of Science and Innovation through the ``Unit of Excellence Mar\'{\i}a de Maeztu
2020-2023'' award to ICCUB. The work of L.M.R. was partly supported by the Spanish
MINECO Grant No.\ PGC2018-094583-B-I00. M.A has been partially supported by the Spanish MINECO Grant No.\ PID2019-104888GB-I00. 
P.B. acknowledges the support from MANF Fellowship of UGC, India.
\end{acknowledgments}

\end{document}